\documentstyle[seceq,epsf,letter]{ptptex}
\newcommand{\beq}{\begin{equation}}
\newcommand{\beqa}{\begin{eqnarray}} 
\newcommand{\eeq}{\end{equation}}
\newcommand{\eeqa}{\end{eqnarray}}
\newcommand{\be}{\begin{equation}}
\newcommand{\ba}{\begin{eqnarray}}
\newcommand{\ee}{\end{equation}}
\newcommand{\ea}{\end{eqnarray}}
\newcommand{\simg}
   {\mathrel{\raise.3ex\hbox{$>$\kern-.75em\lower1ex\hbox{$\sim$}}}}
\newcommand{\siml}
   {\mathrel{\raise.3ex\hbox{$<$\kern-.75em\lower1ex\hbox{$\sim$}}}}

\preprintnumber{
}

\title{
The Luminosity Distance, the Equation of State, and
the Geometry of the Universe
}

\author{
Takeshi {\sc Chiba}$^{1}$\footnote{ JSPS research fellow} and 
Takashi {\sc Nakamura}$^{2}$
}

\inst{
$^{1}$Department of Physics, University 
of Tokyo, Tokyo 113-0033, Japan\\
$^{2}$Yukawa Institute for Theoretical Physics, 
Kyoto University, Kyoto 606-8502, Japan
}

\recdate{
\today
}

\abst{
The second derivative of the luminosity distance with respect to the
redshift is written in terms of the deceleration parameter $q_0$. 
We point out that the third derivative contains the information
regarding the sound speed of cosmic matter as well as the curvature of the
universe. We restrict physically possible parameter ranges of the 
coefficients. It is found that there is a relation between the
coefficients in a flat
universe model with matter such that $c_{s0}(1+w_{\rm x0})=0$  
($c_{s0}$ is the total sound speed of the matter component and
$w_{\rm x0}$=$p_{\rm x0}/\rho_{\rm x0}$). 
}

\begin{document}

\maketitle

\section{Introduction}

There is growing observational evidence that the mean mass density
of the universe is significantly less than the critical
density\cite{pd,os,dbw}. The
cosmological constant is usually introduced to reconcile observations
with the inflationary prediction of a spatially flat universe.
However, the introduction of non-zero cosmological constant
requires the fine-tuning of the vacuum energy, and at present we do not
have any convincing explanation for the reason
why such an extremely small value of the
cosmological constant (in Planck units) is required.

Under these circumstances, several extensions of the standard
$\Omega_{M0}=1$ cold dark matter model have been
proposed\cite{tw,xmatter,qmatter}. Here $\Omega_{M0}$ is the
present ratio of the mean mass density of the universe to the critical
density. Essentially, these models
are characterized by an ``x-component'' having two parameters;
the present ratio of its pressure to its energy density
$w_{\rm x0}=p_{\rm x0}/\rho_{\rm x0}$ and the present sound
speed $c_{s{\rm x0}}^2=\delta p_{\rm x}/\delta\rho_{\rm x}$
\cite{xmatter,qmatter,xmatter2}. Unfortunately, however,
we do not presently have principles regarding the scalar field
potential or the equation of state. Indeed, the situation concerning
the dark matter component
is becoming dark. However, it should be noted that any physically
sensible attempts to include the x-component should introduce
$c_{s{\rm x0}}^2\geq 0$ independent of  $w_{\rm x0}$\cite{qmatter,xmatter2}.

Recently, on the other hand, projects to discover high-redshift
Type Ia supernovae are ongoing, and data are accumulating up to
$z\simeq 1$\cite{sn1,sn2}. These data may indicate that 
the expansion of the universe is accerelating rather than
decerelating\cite{sn3,sn4}. There is another indication for the
existence of an x-component with  $w_{\rm x0} \siml -0.6$\cite{sn5}. 
In this letter, we consider the interpretation of data in more general 
context. To be specific, we attempt to extract useful information 
regarding the equation of state of the universe directly from 
observational data 
rather than to adjust a particular scalar field potential or equation of
state to observations.
The recent observations by the Type Ia supernovae cosmology project are 
providing the information regarding the third derivative (denoted as $d_3$) 
as well as the second derivative ($d_2$) of the luminosity
distance with respect to the redshift\cite{perlmutter}. 
First, we restrict possible parameter ranges of $d_2$ and $d_3$ 
using physical considerations. We then derive a  relation 
between  $d_2$ and $d_3$ for flat models. We discuss the implications of
this consistency relation. 

\section{Luminosity Distance and Equation of State of the Universe}

We assume that our universe can be described by a homogeneous and isotropic
(FRW) model. The luminosity distance is defined by
\beqa
d_L(z)&=&a_0(1+z)f(\chi),\\
\chi&=&{1\over a_0}\int^z_0{du\over H(u)},
\eeqa
where $a_0$ is the present scale factor and $f(\chi)=\chi, 
\sinh (\chi)$ and $\sin (\chi)$ for flat, open
and closed universes, respectively. We expand $d_L(z)$ as 
$H_0d_L(z) = \sum_{i=1}d_i z^i$ and determine $d_2$ and $d_3$ ($d_1=1$ 
by definition). 
$H(z)$ may be expanded as
\beq
H(z)=H_0+{d H \over dz}\bigg{|}_0 z+ {1\over 2}{d^2 H\over
dz^2}\bigg{|}_0z^2 +\dots~~ ,
\eeq
where $``0''$ indicates that the value to be evaluated is that at the 
present epoch. By using the relation
\beq
{dt\over dz}=-{1\over (1+z)H(z)},
\eeq
we obtain
\beqa
{d H \over dz}\bigg{|}_{0}&=&-{\dot H\over (1+z)H}\bigg{|}_{0}=
(1+q_0)H_0,\\
{d^2H\over dz^2}\bigg{|}_{0}&=& {\ddot H\over
(1+z)^2H^2}\bigg{|}_{0}+\dot H\left(
{1\over (1+z)^2H}-{\dot
H\over(1+z)^2H^3}\right)\bigg{|}_{0}\nonumber\\
&=&(c_0+3q_0+2)H_0-(2+3q_0+q_0^2)H_0,
\eeqa
where $q_0$ and $c_0$ are defined by\footnote{``$c$'' stands
for cubic, not to be confused with the sound speed, $c_s$.}
\beqa
q_0&\equiv & -{\ddot a a\over \dot a^2}\bigg{|}_{0},\\
c_0&\equiv & { a^{\cdots} a^2\over \dot a^3}\bigg{|}_{0}.
\eeqa
Then we have
\beq
H_0d_L(z) = \sum_{i=1}d_i z^i
= z+{1\over 2}(1-q_0)z^2+{1\over 3!}(3q_0^2+q_0-c_0-\Omega_{k0}-1)z^3 
+\dots,
\label{expansion}
\eeq
where $\Omega_{k0}\equiv k/(a_0H_0)^2$ with $k=0, -1$ and  $1$ for 
flat, open and closed universes, respectively.

Thus $d_2$ and $d_3$ are written as
\beqa
d_2&=&{1\over 2}(1-q_0)\\
d_3&=& {1\over 6}(3q_0^2+q_0-c_0-\Omega_{k0}-1)
\eeqa
We find that $d_3$ contains information concerning $c_0$ and the curvature
of the universe. The meaning of $c_0$ will be clarified later. 

We have not yet specified the properties of the cosmic matter. 
Now we restrict the possible ranges of $d_2$ and $d_3$ on the basis of 
physical requirements. We impose the following physically reasonable 
requirements on the cosmic 
matter: (1) the total density is non-negative; (2) the total 
density is  currently not increasing as 
a function of time; (3) the present sound speed $c_{s0}$ of the total
system  satisfies  
$0\leq c_{s0}^2 \leq 1$ for causality and local stability\cite{xmatter}.
The first and second requirements are equivalent to assuming the weak 
energy condition\cite{he}.\footnote{In a $\Lambda <0$ dominated
universe, the first condition can be violated. We exclude such a case.}

The first requirement implies 
\beq
\Omega_{k0}\geq -1.
\label{first}
\eeq
By  use of the relation
\beq
\dot{\rho}=-3H(\rho+p)={3H\over 4\pi}\left(\dot{H}-{k\over a^2}\right),
\eeq
we find that the second requirement implies
\beq
1+q_0+\Omega_{k0}\geq 0.
\label{second}
\eeq
The sound speed of the total system, $c_{s0}$, can be written as
\beq
c_{s0}^2={\dot{p}\over \dot{\rho}}\bigg{|}_0=
{c_0-\Omega_{k0}-1\over 3(1+q_{0}+\Omega_{k0})}.
\eeq
Then the third requirement, together with Eq.(\ref{second}), implies
\beq
1+\Omega_{k0} \leq c_0 \leq  4(1+\Omega_{k0}) +3q_0.
\label{third}
\eeq

Now we rewrite these constraints in terms of $d_2$, $d_3$ and
$\Omega_{k0}$. The second constraint is rewritten as
\beq
d_2 \leq 1+ {1\over 2}\Omega_{k0}.
\label{a2}
\eeq
The third constraint is rewritten as
\beq
2d_2^2-{4\over 3}d_2 -{2\over 3} -{5\over 6}\Omega_{k0}\leq d_3
\leq 2d_2^2-{7\over 3}d_2 +{1\over 3} -{1\over 3}\Omega_{k0}.
\label{a3}
\eeq
To conclude, the region bounded by the relations in Eq.(\ref{first}) 
and Eq.(\ref{a2}) and 
Eq.(\ref{a3}) is allowed by three requirements.  
The allowed region is shown in Fig.1 for several $\Omega_{k0}$. 


The quantities $q_0$ and $c_0$ can also be written in terms of 
the dark matter components as
\beqa
q_0&=& \sum_i{\Omega_{i0}\over 2}\left(1+3w_{i0}\right),\\
c_0&=& 1+\Omega_{k0}+{9\over 2}c_{s0}^2\sum_i\Omega_{i0}(1+w_{i0}),
\eeqa
where $i$ denotes the $i$-th  component, and  
$w_{i0}=p_{i0}/\rho_{i0}$. We also note that the total sound speed 
$c_{s0}^2$ can be written 
as
\beq
c_{s0}^2={\sum_i c_{si0}^2\Omega_{i0}(1+w_{i0})
\over \sum_i \Omega_{i0}(1+w_{i0})}.
\eeq
Thus, we see that $c_0$ (or $d_3$)
contains  information regarding the sound speed of the dark matter
component.  In the xCDM model (that is,
CDM plus x-component; x can be $\Lambda$ or anything else), 
$q_0$ and $c_0$  can be written as 
\beqa
q_0&=& {\Omega_{M0}\over 2}+{\Omega_{\rm x0}\over 2}\left(1+
3w_{\rm x0}\right),\\
c_0&=& 1+\Omega_{k0}+{9\over 2}c_{s0}^2\Omega_{\rm x0}(1+w_{\rm x0}).
\eeqa
For example, consider x to be $\Lambda$ and take 
$(\Omega_{M0},\Omega_{\Lambda 0})=(0.40,0.70)$, 
that is, $(q_0,c_0)=(-0.50,1.10)$ so that the universe is closed. 
However, in  the xCDM model, even a flat model with
$(\Omega_{\rm x0},w_{\rm x0},c_{s0}^2)\simeq(0.80,-0.83,0.33)$ is allowed. 
The point is that the values of $d_2$ and $d_3$ can not determine the
equation of state of the universe uniquely. Depending on  
the values of $d_2$ and $d_3$, an open or closed $\Lambda$CDM model
may be interpreted by a flat xCDM model, for example.
This degeneracy can be resolved only when $\Omega_{k0}$ is determined
independently, which may be done by observing the variation of 
the redshift in the absorption lines of quasars\cite{loeb}.

We would like to point out that there is an interesting
 ``consistency relation'' in flat models: If we assume
$c_{s0}(1+w_{\rm x0})=0$  and a flat model, then $d_3$ 
can be written in terms of $d_2$ as 
\beqa
d_2&=&{1\over 2}(1-q_0),\\
d_3&=& {1\over 6}(3q_0^2+q_0-2)=2d_2^2-{7\over 3}d_2+{1\over 3}. 
\label{consist}
\eeqa
Although this relation seems rather trivial, the contraposition of the 
above statement has important meanings. Namely,  {\it if the
consistency relation} Eq.({\ref{consist}) {\it 
did not hold in the observational data, 
then it would follow that the dark matter component has the 
property} $c_{s0}(1+w_{\rm x0}) \neq 0$, {\it or the Universe is not flat}. 
Note that for $\Lambda$-dominated universe
models, the former possibility is automatically satisfied. 
 The consistency relation Eq.(\ref{consist}) is 
represented in Fig.1 for $\Omega_{k0}=0$ as a thick curve.
Of course, the current observational data suffer from uncertainties,
and it is premature to put the consistency relation into
practice. However, if observations  become sufficiently accurate, we
should prepare ourselves for the possibility that the dark matter 
component has the property $c_{s0}(1+w_{\rm x0}) \neq 0$, 
{\it or} the Universe is not flat. Both possibilities should be of
great significance.

\section{Conclusion}

We have shown that the third derivative of the luminosity distance
with respect to redshift contains information regarding the sound speed 
of the cosmic matter as well as the curvature of the universe. 
We have also studied the possible parameter ranges of the expansion
coefficients of the luminosity distance and have derived a consistency 
relation for the flat universe model with the cosmic matter of
$c_{s0}(1+w_{\rm x0}) = 0$. 
The number of data taken by the supernova cosmology project is
increasing, and better statistics will be available in future.
Then the more general possibility presented in this paper
may be the case. 

\vspace{3mm}

We would like to thank Peter Garnavich for drawing our 
attention to Ref.\cite{sn5}. 
One of the authors (TC) is supported by JSPS Research Fellowship for 
Young Scientists. This work was  supported by a Grant-in-Aid  of the
Ministry of Education, Culture, and Sports No.09640351.

\vskip 2cm

\section*{Figure Caption}

\vspace*{12pt}
\noindent
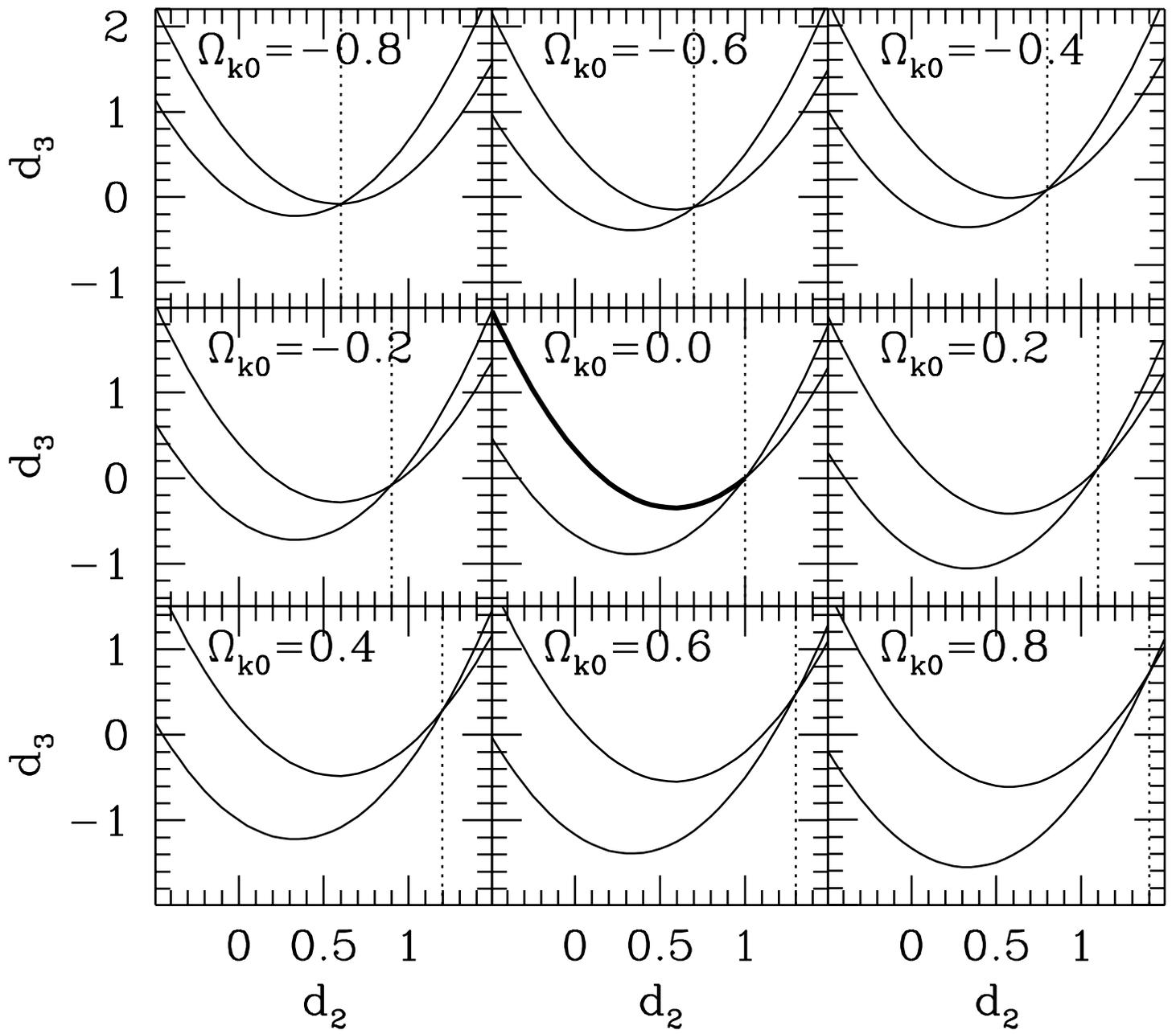
{\bf Figure 1: }
The allowed regions of $d_2$ and $d_3$ for several 
$\Omega_{k0}$.
A region that is bounded by two curves 
and is on the left-hand-side of the vertical line is allowed. 
For $\Omega_{k0}=0$, the consistency relation is indicated by the
thick curve.

\end{document}